\documentclass[12pt, aaspp4, flushrt, mathptmx]{aastex}
\usepackage{epsf}
\usepackage{verbatim}

\newcommand{\comments}[1]{}

\begin{document}

\sffamily

\title{Fractal multi-scale nature of solar/stellar magnetic field}

\author{Valentina I. Abramenko}

\affil{\it Big Bear Solar Observatory, New Jersey Institute of Technology, Big Bear City, CA 92314, USA}

\begin{abstract}
An abstract mathematical concept of fractal organization of certain complex
objects received significant attention in astrophysics during last decades. The
concept evolved into a broad field including multi-fractality and intermittency,
percolation theory, self-organized criticality, theory of catastrophes, etc.
Such a strong mathematical and physical approach provide new possibilities for
exploring various aspects of astrophysics. In particular, in the solar and
stellar magnetism, multi-fractal properties of magnetized plasma turned to be
useful for understanding burst-like dynamics of energy release events,
conditions for turbulent dynamo action, nature of turbulent magnetic
diffusivity, and even the dual nature of solar dynamo. In this talk, I will
briefly outline how the ideas of multi-fractality are used to explore the above
mentioned aspects of solar magnetism.

\end{abstract}

\comments{turbulence, Sun: magnetic fields, Sun: granulation, Sun: photosphere, 
instrumentation: high angular resolution, methods: data analysis}

\section{Introduction: Why Fractals?}

\noindent A mathematical fractal is a self-similar object on all possible
spatial and time
scales. It means that when we proceed from large to smaller scales, we will see
exactly the same picture. From mathematical standpoint this means that a unique
scaling law holds for all scales. A fractal (or, more rigorously, a
mono-fractal) is a deterministic, predictable system. Mathematical mono-fractals
differ drastically from what we observe in nature: fractal-like structures in
nature are  \textit {multi-fractals} -  a superposition of infinite number of
mono-fractals. 

The transition from mono-fractals to multi-fractals turns an amusing
mathematical toy into a powerful tool to study real processes in nature. The
matter is that multi-fractals posses the same properties in both the spatial and
temporal domains. This means that if we see a very complex, jagged shape in
space (multi-fractal in space), then we will observe a violent, burst-like
behavior in time (multi-fractal in time). For such systems, any small
perturbation can cause an avalanche of any possible size. 

Therefore, revealing the fact that a system under study is a fractal does not
allow us to make inferences about its nature and essential properties of its
behavior. We need something more, namely, to know of how many mono-fractals our
system is made of. Numerous examples of fractals and multi-fractals can be found
 \citep[e.g.,][Internet]{feder1989,Schroeder2000}.

Mathematical details of the
fractal calculus are well described as
well \citep[e.g.,][]{Baumann2005,McAteer+2007}.
Historically, when analyzing spatial objects, their capability to be organized
into very jagged structures with extended voids and sharp peaks is addressed as
a property of multi-fractality. At the sate time, while analyzing time series,
we address the same property as intermittency. Thus, multi-fractality and
intermittency are two terms for the same physical property of a system. I will
use both of them in this talk.

Several approaches were elaborated during a couple of last decades to probe the
properties of multi-fractality in different fields of science. In a brief review
below, I will focus on multi-fractality techniques applied in astrophysics.
Thus, multi-fractal systems are capable of self-organization (i.e., formation of
larger entities from smaller ones via inverse cascade), and of self-organized
criticality (SOC) when burst-like energy release events of any scale are
possible at any moment
\citep[e.g.,][]{Charbonneau+2001,Longcope-Noonan2000,Aschwanden2011a-SolPhys}.

 A theory of catastrophes is also
based on the multi-fractal nature of astrophysical phenomena
\citep{Priest-Forbes2002,Isenberg-Forbes2007}.
Percolating clusters
\citep{Balke+1993,Seiden-Wentzel1996,Pustilnik-1999,Schatten-2007}
 are also fractals and multi-fractals.

Direct calculations of fractal dimensions and spectra of multi-fractality is one
of the most popular tool to explore astrophysical multi-fractals 
\citep[e.g.,][]{lawrence1993,Meunier-1999,Lepreti+1999,McAteer+2007,
Dimitropoulou+2009,Abramenko_Int_2010,Acshwanden_2011}.

Another possibility to study multi-fractality is to analyze high statistical
moments by means of distribution functions \citep{Bogdan+1988,
Parnell_2009}, or
structure functions \citep{Consolini+1999,Abramenko+2002,abramenko-int-2005,
Uritsky+2007,abramenko+yurchyshyn-int-2010,Abramenko+2012-mini}.

{\bf Essential physical properties of multi-fractals } can be formulated as
follows. 

(i) Scaling laws change with scale, i.e., no unique power law index can be valid
for all scales; 

(ii) Large fluctuations (in both time and space domains) are not rare
and contribute significantly to high statistical moments, which grow as the data
set expands;

(iii) Direct and inverse cascades along scales are possible (fragmentation and
aggregation), which results in capability to form larger features from smaller
ones, i.e., self-organization and SOC state.
 
These properties can help us to diagnose the presence and degree of
multi-fractality of various astrophysical phenomena. Meanwhile, keeping in mind
that in astrophysical magnetism we deal with a specific type of a multi-fractal
medium, namely, \textit {intermittent turbulence in an electro-conductive flow},
we can take advantage of it and incorporate other very important properties and
tools. So our list of properties can be extended:

(iv) Intermittent turbulent magnetized plasma is capable of amplifying a seed
magnetic field, i.e., local fast dynamo is at work 
\citep{Zeldovich+1987-UFN,biskamp1993,Vogler-Schussler-2007-SSD,
PietarilaGraham_2010}, 
see also a recent review by \citet{Brandenburg+2012review};

(v) Turbulent plasma at high Reynolds number displays properties of
multi-fractality and intermittency in spatial/temporal structures of
temperatures, velocities, density, etc. 
\citep[see, e.g.,][]{Zeldovich+1987-UFN,frisch1995};

(vi) The regime of diffusivity on multi-fractals is expected to be an anomalous
diffusion. 

Based on these properties of multi-fractal systems, I will discuss below how
exploration of these properties can help us in understanding of solar and
stellar magnetism. It is impossible in the framework of this invited talk to
discuss all aforementioned approaches and tools in great details, so I will
concentrate on the analysis of structure functions, which are used in my
research.

\section{Structure functions approach to study multi-fractality}

\begin{figure}[p]
\centering
\begin{tabular}{c}
\epsfxsize=6.0truein  \epsffile{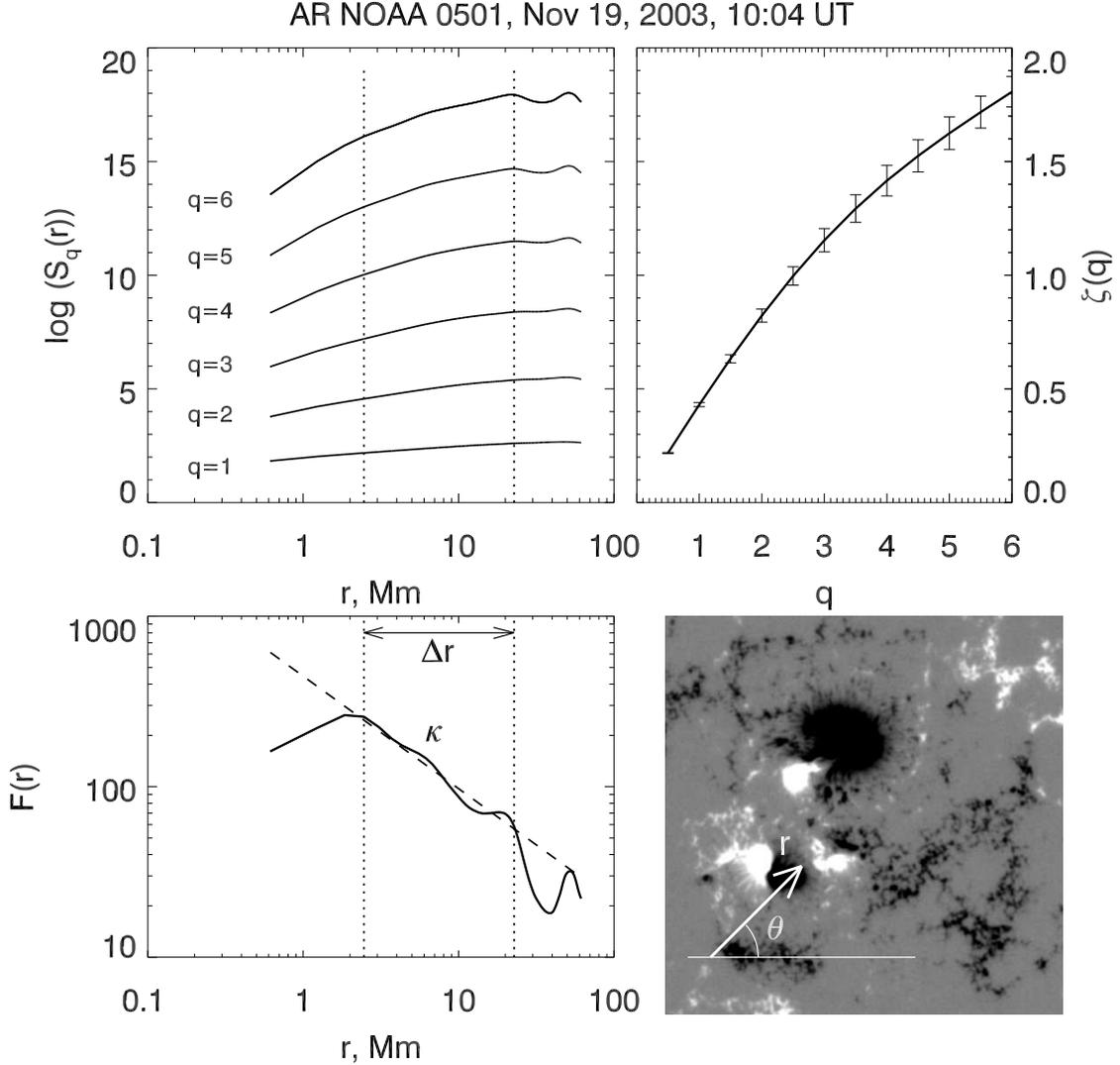} \
\end{tabular}
\caption{Structure functions $S_q(r)$ ({\it upper left}) calculated from a
magnetogram of active region NOAA 0501 ({\it lower right}) according to Eq.
(\ref{Sq}). {\it Lower left} - flatness function $F(r)$ derived from the
structure functions using Eq. (\ref{Fr}). Vertical dotted lines in both
left panels mark the interval of intermittency, $\Delta r$, where flatness grows
as power law when $r$ decreases. The index $\kappa$ is the power index of the
flatness function determined within $\Delta r$. The slopes of $S_q(r)$, defined
for each $q$ within $\Delta r$, constitute $\zeta(q)$ function ({\it upper
right}), which is concave (straight) for a multi-fractal/intermittent
(mono-fractal/non-intermittent) field. An example of a separation vector $\bf{
r}$ and the corresponding directional angle $\theta$  are shown on
the magnetogram.}
\label{fig1}
\end{figure}

\noindent Since Kolmogorov's study \citep{Kolmogorov_1941},
various
models have been proposed to describe the statistical behavior of fully
developed turbulence. In these studies, the flow is modeled using statistically
averaged quantities, and structure functions play a significant role. They are
defined as statistical moments of the $q-$powers of the increment of a field.
The definition can be applied to different fields (e.g., velocity, temperature,
magnetic field, etc). Here, in the most of the cases, I will refer to the
line-of-sight component of the magnetic field,  $B_l$, for which the structure
function can be written as 
\begin{equation}
S_q(r) = \langle | {B_l}({\bf x} + {\bf r}) - {B_l}({\bf x})|^q \rangle, 
\label{Sq} 
\end{equation}  
where ${\bf x}$ is the current pixel on a magnetogram,  ${\bf r}$ is the
separation vector between any two points used to measure the increment (see the
lower right panel in Figure \ref{fig1}, and $q$ is the order of a statistical
moment, which takes on real values. The angular brackets denote averaging over
the magnetogram, and the vector $\bf r$ is allowed to adopt all possible
orientations, $\theta$,  on the magnetogram. The next step is to calculate the
scaling of the structure functions, which is defined as the slope, $\zeta(q)$,
measured inside some range of scales where the $S_q(r)$-function is linear and
the field is intermittent. The function $\zeta(q)$ is shown in the upper
right panel in Figure \ref{fig1}.

%#####################################################################
%#####################################################################

A weak point in the above technique is the determination of the range, $\Delta
r$, where the slopes of the structure functions are to be calculated. To
visualize the range of intermittency, $\Delta r$, we suggest to use the flatness
function \citep{abramenko-int-2005}, which is determined as the  ratio of the
fourth
statistical moment to the square of the second statistical moment. To better
identify the effect of intermittency, we reinforced the definition of the
flatness function and calculated the hyper-flatness function, namely, the ratio
of the sixth moment to the cube of the second moment:
\begin{equation} 
F(r)=S_6(r)/(S_2(r))^3 \sim k^{ - \kappa}.
\label{Fr}
\end{equation} 
For simplicity, we will refer  to $F(r)$ as the flatness function, or
multi-fractality/intermittency spectrum. For a non-intermittent  structure, the
flatness function is not dependent on the scale, $r$. On the contrary, for an
intermittent/multi-fractal structure, the flatness grows as power-law, when the
scale decreases. The slope of flatness function, $\kappa$, and the width of
$\Delta r$ characterize the degree of multi-fractality and intermittency.

Application of this technique to two hundred of solar active regions observed
with SOHO/MDI in the high resolution mode \citep{Abramenko_Int_2010}
demonstrated that active regions of high flare productivity display steeper and
broader multifractality spectra, $F(r)$. The inference agrees with the
formulated above
statement that multi-fractality in spatial domain is accompanied by
intermittency (burst-like behavior) in time. Moreover, for any multi-fractal
system, individual bursts cannot be precisely predicted in advance. So, the
exact prediction of the location and the onset moment of a flare (of any size),
strictly speaking, is a hopeless task. Based on different indirect indications,
one may only hope to provide a probabilistic estimate for ongoing flaring.

Multi-fractality of time series of X-ray emission from an individual solar
flare was discussed in \citet{McAteer+2007}, where
an inference on the fractal nature of the flaring current sheet was made.  

I will focus below on a solar surface outside active regions, which occupy
usually more than 80\% of the entire solar surface. Many important aspects
of solar magnetism are rooted there. 

\section{Multi-fractality in the solar surface}

\begin{figure}[h]
\centering
\begin{tabular}{c}
\epsfxsize=6.2truein  \epsffile{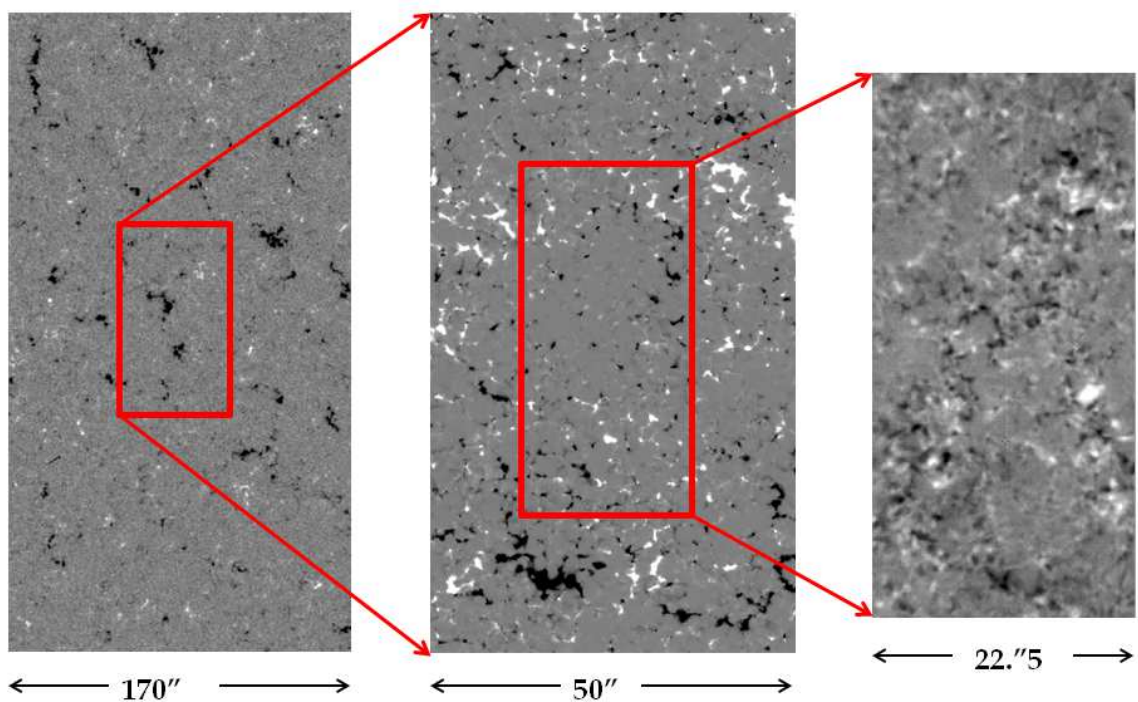} \
\end{tabular}
\caption{Examples of LOS magnetograms recorded inside CHs with three
solar instruments (from left to right): SDO/HMI magnetogram (Aug 12, 2011,
spatial sampling of 0.$''$5); SOT/SP magnetogram (Mar 10, 2007, spatial sampling
of 0.$''$16); BBSO/NST magnetogram (Jun 2, 2012, spatial sampling of 0.$''$098).
Red boxes and arrows outline areas of the same size.}
\label{fig2}
\end{figure}

Examples of line-of-sight (LOS) magnetograms recorded in coronal holes (CHs)
using different solar instruments are shown in Figure \ref{fig2}. The left panel
shows data from the Helioseismic and Magnetic Imager (HMI) onboard the Solar
Dynamics Observatory \citep[SDO, ][]{Scherrer+2012-SDO}, the
middle panel shows LOS magnetic field provided by Hinode Solar Optical Telescope
Spectro-Polarimeter \citep[SOT/SP, ][]{tsuneta2008},
and
the right panel presents a magnetogram from the New Solar Telescope (NST) 
\citep{Goode+NST-2010} operating at the Big Bear Solar
Observatory (BBSO). Note that the NST data were obtained at a near-infrared
spectral line (1.56 $\mu m$) and represent the magnetic field in the deep
photosphere at depths of about 50~km below the $\tau_{500}=1$ level.
Figure \ref{fig2} clearly demonstrates that with improved telescope resolution
more mixed polarity magnetic elements become visible inside a CH. In spite of
the fact that the three magnetograms refer to different CHs, this tendency is
well defined.
%#####################################################################
%#####################################################################

\begin{figure}[h]
\centering
\begin{tabular}{c}
\epsfxsize=5.3truein  \epsffile{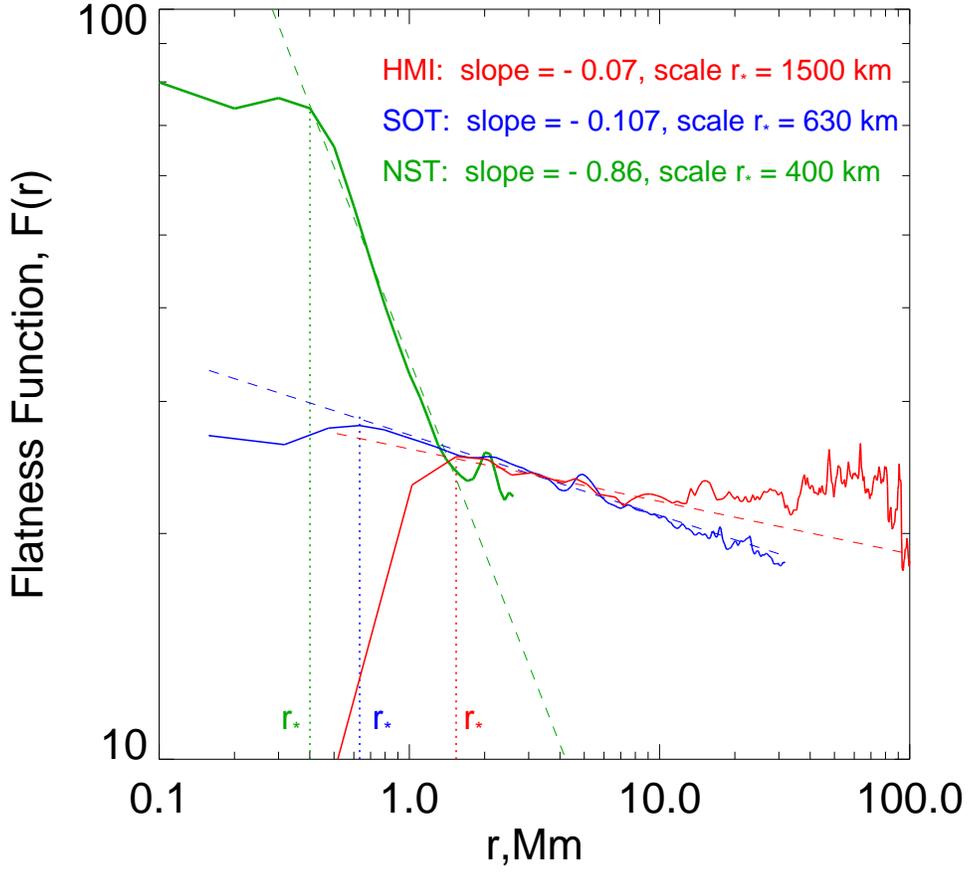} \
\end{tabular}
\caption{Flatness functions calculated from the magnetograms shown in
Fig.\ref{fig2}. 
Dashed lines show best linear fits to the data points inside intervals starting 
above the small-scale cutoff, $r_{\star}$. For better compatison, the curves
are shifted along the vertical axis.}
\label{fig3}
\end{figure}

{\underline{\it Flatness functions}} calculated from the three above mentioned
magnetograms are shown in Figure \ref{fig3}. The HMI data show only a hint of
multi-fractality on scales above 1500~km and a very shallow slope of $F(r)$
($\kappa =-0.07$). The HMI resolution of 1$''$ obviously is not sufficient to
clearly reveal multi-fractality in quiet Sun. Meanwhile, the SOT/SP data show a
much broader scale range of multi-fractality down to approximately 630~km and a
steeper slope ($\kappa =-0.107$). The HMI result refers to the height of
280-360~km (the effective line formation level of the FeI 617.3 nm spectral
line, \citet{Gortovenko-Kostyk-1989}),
 whereas the SOT/NBF
data refer to a level of 400-700~km in the photosphere (the line formation
height of the NaI 589.6 nm spectral line is discussed in 
\citet{Sheminova-1998}). The flatness function obtained from NST data clearly
reveal (at the deeper layer) strong intermittency and multi-fractality on scales
down to $\sim$400~km.
%#####################################################################
%#####################################################################

Thus, the multi-fractal nature of small-scale magnetic fields becomes better
pronounced with depth and improvement of spatial resolution, which leads 
us to conclude that intermittency and multi-fractality is an intrinsic property
of the near-surface magnetic fields in the quiet Sun.

\begin{figure}[h]
\centering
\begin{tabular}{c}
\epsfxsize=5.3truein  \epsffile{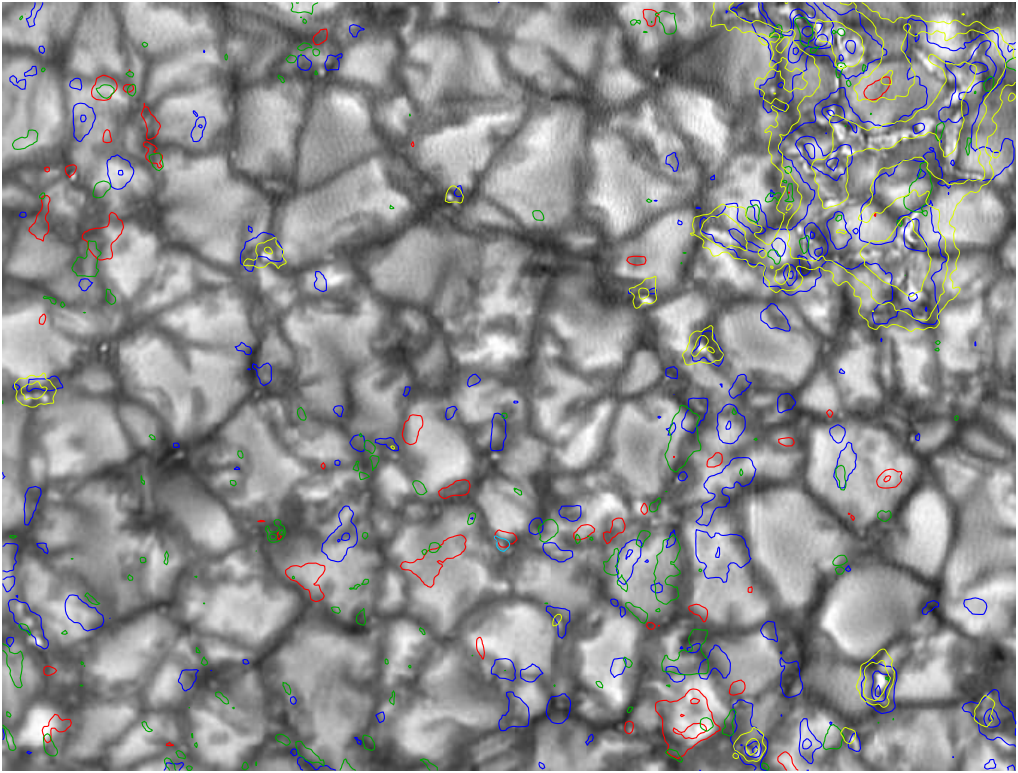} \
\end{tabular}
\caption{Background - NST/TiO image of 26.6$''$0 $\times$ 20.1$''$ in size
recorded in a CH at 18:20:32 UT on Aug 12, 2011. Blue (red) contours show the
line-of-sight positive (negative) component measured with NST and correspond to
90, 210, 300 G. Green contours represent the signal from the transverse magnetic
field component from NST, $(Q^2+U^2)^{1/2}$, corresponding to 100 and 200 G.
Yellow and turquoise contours outline the Hinode SOT/NBF magnetic elements of
negative (-50, -100, -300 G) and positive (50 G) polarities
for the same day and time.}
\label{fig4}
\end{figure}

{\underline{\it Magnetic elements against granulation}} inside a CH are
shown in Figure \ref{fig4}, where the background is an NST solar granulation
image overplotted with NST LOS and transverse magnetic features and
Hinode SOT Narrow Band Filter (NBF) LOS magnetic field.

%#####################################################################
%#####################################################################

In the upper right corner of the image, a fragment of bright points (BPs)
filigree corresponding to the super-granular boundary is visible. The rest of
the image shows the intra-network area, where nine isolated magnetic elements
were detected by the SOT/NBF. All of them are co-spatial with BPs and with the
LOS signal from the NST. In six cases there are neither opposite polarity nor
transverse magnetic field features in the closest vicinity. This may indicate
that these magnetic elements are footpoints of open magnetic flux tubes
representing the skeleton of the CH. The presence of BPs indicates that they
might be produced via the convective collapse \citep{Parker-1978,Spruit-1979}.

At the same time, a significant part of the intra-network population is composed
of magnetic features, which are not related to BPs and scattered over granules
and inter-granular lanes. These magnetic elements were not detected by SOT/NBF
and they are not necessarily very weak. On the contrary, they are quite
compatible (by size and intensity) with those detected by the both instruments,
they are simply not visible in NBF magnetograms. A simplest explanation could be
the difference in heights. As I mentioned above, the NST measures the magnetic
signal formed very deep in the photosphere, precisely, at the depth of 50 km
below the $\tau_{500} = 1$\ level, while the magnetic signal measured with
SOT/NBF is formed at the hight of approximately 400-700 km. If magnetic elements
associated with granules are predominantly small (200-500 km in length) closed
loops anchored at a depth of about -50 km, they might not be visible at the
altitude of 400-700 km. This speculation is supported by the inspection of
mutual location of Stokes V features (blue and red contours) and the transverse
magnetic field features (green contours) in Figure \ref{fig4}. Indeed, the
V-signal is co-spatial with the $(Q^2 + U^2)$ signal for the granules-associated
magnetic elements, which supports the idea of closed loops (or bunches of loops)
rather than presence of singular footpoints of extended, high loops or open
field lines. 

As for the magnetic elements associated with BPs and visible with the both
instruments, they seem to be the best candidates for the roots of 
the open field lines, as we mentioned above. This can explain why they are
visible on different
heights.

Thus, the data allows us to speculate that the BPs-associated magnetic elements
are related to the advection and convective collapse, whereas the numerous
granule-associated intra-network elements are situated deeper in the photosphere
and might be produced (at least, part of them) by local turbulent dynamo (see
the talk by Dr. Tsuneta in this Symposium).
%==========================================================
%$$$$$$$$$$$$$$$$$$$$$$$$$$$$$$$$$$$$$$$$$$$$$$$$$$$$$$$$$$
%==========================================================

\begin{figure}[h]
\centering
\begin{tabular}{c}
\epsfxsize=6.5truein  \epsffile{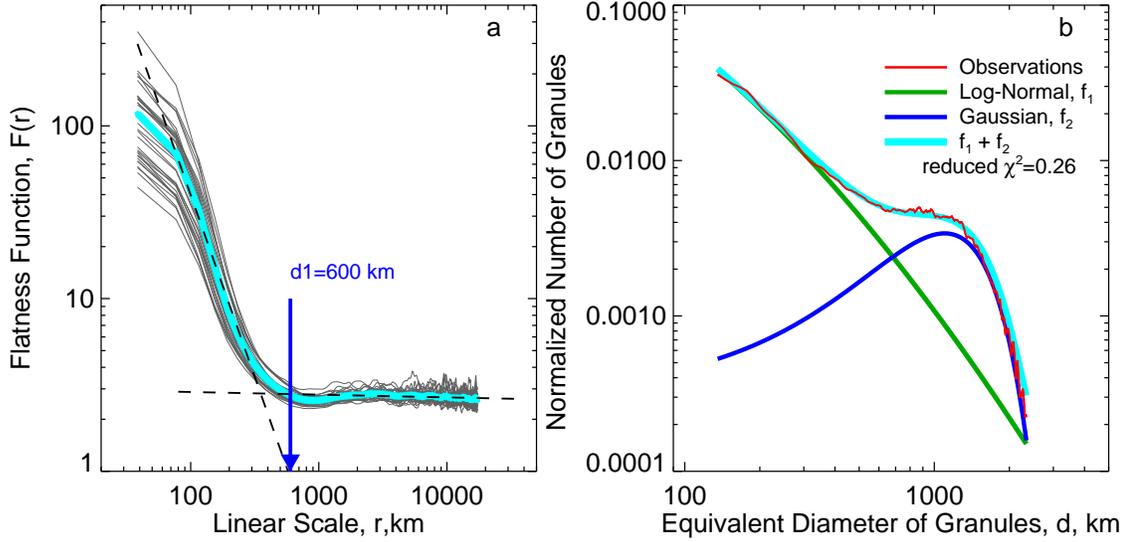} \
\end{tabular}
\caption{{\it a} - Flatness functions calculated from 36 granulation images
(gray) and their average (turquoise). The dashed segments show the best linear
fits to the data points. The blue arrow divides the multi-fractality range where
the flatness function varies as a power law from the Gaussian range where the
flatness function is scale independent. {\it b} - decomposition of the observed
averaged probability distribution function (red line) into two components: a
log-normal approximation, ($f_1$, green line) and a Gaussian approximation
($f_2$, blue). Their sum is plotted with the turquoise line.}
\label{fig5}
\end{figure}

{\underline{\it Multi-fractality of granulation}}. To drive local turbulent
dynamo, the environment should be a highly turbulent medium, i.e., to be a
multi-fractal.  Is the solar granulation pattern a multi-fractal? To explore the
question, \citet{Abramenko+2012-mini}
 used a NST data set
of solar granulation images obtained for the quiet Sun area on the solar disk
center recorded under excellent seeing conditions. Flatness functions for 36
independent snapshots and their average are shown in Figure \ref{fig5}{\it a}.
%#####################################################################
%#####################################################################

The flatness functions indicate that solar granulation is non-intermittent (a
mono-fractal) on scales exceeding approximately 600 km, and it becomes highly
intermittent and multi-fractal on scales below  600 km. Thus, a random,
Gaussian-like distribution of granule size holds down to 600 km only. On smaller
scales, the multi-fractal spatial organization of solar granulation takes over. 

A distribution function of granular size (Figure \ref{fig5}{\it b}) further
confirms this inference. On scales of approximately 600 and 1300 km, the
averaged probability distribution function (PDF) rapidly changes its slope. This
varying power law PDF is suggestive that the observed ensemble of granules may
consist of two populations with distinct properties: regular granules and
mini-granules. Decomposition of the observed PDF showed that the best fit is
achieved with a combination of a log-normal function, $f_1$, representing
mini-granules, and a Gaussian function, $f_2$, representing regular granules.
Their sum perfectly fits to the observational data. 

Until now it was thought that solar convection produces convection cells,
visible on the solar surface as granules, of characteristic ("dominant") spatial
scale of about 1000 km and a Gaussian (normal) distribution of granule sizes. In
this case, the mechanism that produces granules is "programmed" to churn up
convection cells of a typical size, without much freedom in size variation.
Mini-granules do not display any characteristic ("dominant") scale, their size
distribution is continuous and can be described by a decreasing log-normal
(Gaussian distribution does not work any longer here). A majority (about 80\%)
of mini-granules are smaller than 600 km and about 50\% are smaller than 300 km
in diameter. This non-Gaussian distribution of sizes implies that a much more
sophisticated mechanism, with much more degrees of freedom may be at work, where
any very small fluctuation in density, pressure, velocity and magnetic fields
may have significant impact and affect the resulting dynamics. Physical
differences between the log-normal and Gaussian distributions are discussed by,
e.g., \citet{Abramenko-Longcope-2005}.

An important inference from the above discussion reads that a
necessary condition for the seed magnetic field to be amplified is met. So, 
local turbulent dynamo in the near-surface layer is quite a possibility.

%==========================================================
%$$$$$$$$$$$$$$$$$$$$$$$$$$$$$$$$$$$$$$$$$$$$$$$$$$$$$$$$$$
%==========================================================

{\underline{\it Regime of turbulent magnetic diffusion in the photosphere}}. As
we saw in Introduction, the anomalous diffusivity is another hallmark of
multi-fractality. The dispersal process embedded in a multi-fractal cannot
follow the random walk with normal diffusion. For example, in the case of solar
photosphere, the multi-fractal plasma cannot ensure an arbitrary displacement in
an arbitrary direction for all magnetic elements. A discussion of differences
between the normal and anomalous diffusion can be found, e.g., in 
\citet{Lawrence_Schrijver_1993,Vlahos+2008}.

The coefficient of magnetic diffusivity is an essential input parameter
for meridional flux transport models and global dynamo models. Therefore,
magnetic flux dispersal on the solar surface was studies extensively 
\citep[e.g.,][]{Lawrence_Schrijver_1993,Schrijver_1996,
Berger_1998_495,Berger_1998_506,Cadavid_1999,
Hagenaar_1999,Lawrence_2001,Utz_2009,Utz_2010,Crockett_2010,
Abramenko+2011-Diff}.

In the most of these studies, observational data were interpreted in the
framework of normal diffusion, and variety of estimates for the magnetic
diffusivity coefficient, $\eta$, were reported: from 50 km$^2$s$^{-1}$
\citep{Berger_1998_506} to 350 km$^2$ s$^{-1}$ 
\citep{Utz_2010}.
 Numerical simulations of the isotropic turbulence
with magnetic field \citep{Brandenburg+2008}
 showed that the turbulent magnetic diffusivity
increases with increasing scale. Combination of MHD modeling with observations
allowed \citet{Chae_2008}
to conclude that the
turbulent diffusivity changes with scale and is smallest (about 1 km$^2$
s$^{-1}$) on smallest available scale of approximately 200 km.

\begin{figure}[h]
\centering
\begin{tabular}{c}
\epsfxsize=6.5truein  \epsffile{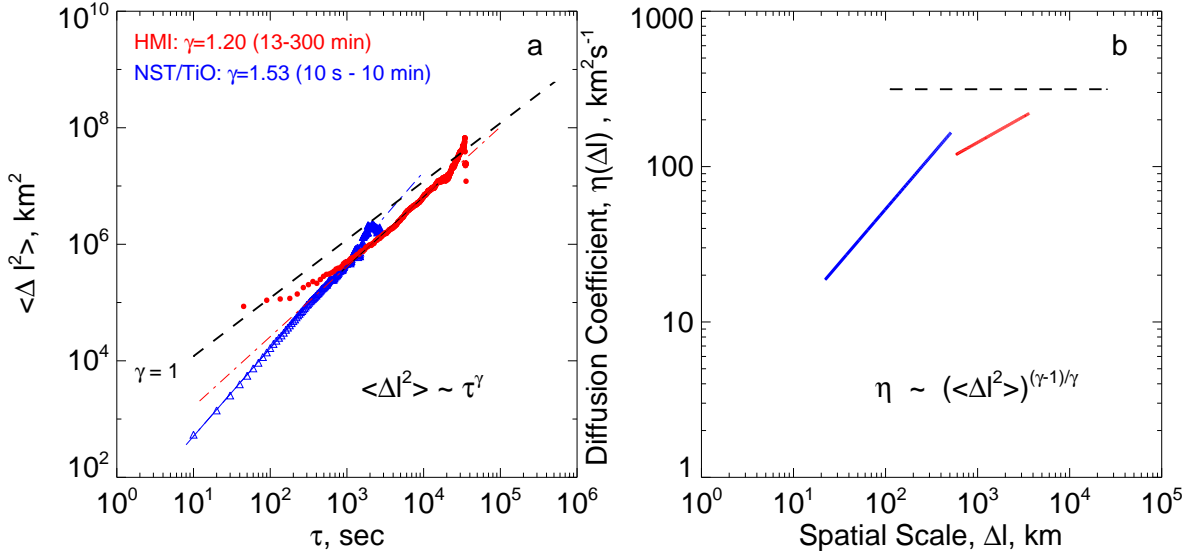} \
\end{tabular}
\caption{{\it a} - Squared displacements of magnetic BPs detected from a 2-hour
data set from the NST/BBSO ({\it blue}) and squared displacements of magnetic
elements detected from 9-hour data set from SDO/HMI magnetograms recorded  in a
quiet Sun area on the solar disk center ({\it red}). Dash-dot lines are the
best linear fit to the data points inside ranges of linearity; the slopes of
the fits, $\gamma$, are indicated. 
{\it b} - The turbulent magnetic diffusivity, $\eta$, as a function of linear
scale derived by Eq. (\ref{K_l}) from linear fits for the NST (blue) and HMI
(red) data shown in panel {\it a}. The thick dashed lines in both panels show
an example of scaling for the normal diffusion regime with $\gamma=1$.}
\label{fig6}
\end{figure}

Photospheric BPs, as tracers of kilo-gauss magnetic flux tubes, were utilized
to probe photospheric flux dispersal \citep[e.g.,][]{Berger_1998_495,
Berger_1998_506,Utz_2010,Crockett_2010}.
Recently, the high resolution power of the NST allowed 
\citet{Abramenko+2011-Diff}
 to explore the regime of diffusion in the
photosphere down to scales of 10 sec in time and 25 km in space. Magnetic BPs
detected from NST/TiO images were tracked, and their squared displacements (from
the initial position of a given BP) were calculated as a function of a time lag,
$\tau$. Later, the routine was repeated for HMI magnetic flux concentrations in
a quiet Sun area on the disk center. Figure \ref{fig6}{\it a} summarizes
results. 

%#####################################################################
%#####################################################################

Recall that for normal diffusion (Brownian motions), the squared displacements
of tracers are directly proportional to time, i.e., the power law index,
$\gamma$, of the displacement spectrum is a unity (an example of the normal
diffusion regime is illustrated in Figure \ref{fig6} with thick black dashed
lines). When $\gamma > 1$ ($\gamma < 1$), a regime of super-diffusion
(sub-diffusion) dominates. The squared displacements $(\Delta l)^{2}(\tau)$ can
be approximated, at a given range of scales, as
\begin{equation}
(\Delta l)^{2}(\tau)=c\tau^{\gamma},
\label{gen}
\end{equation}
where $c=10^{y_{sect}}$ and $\gamma$ and $y_{sect}$ are derived from the
best linear lit to the data points plotted in a double-logarithmic plot.
Then the diffusion coefficient can be written as 
\citep{Abramenko+2011-Diff}:
\begin{equation}
\eta (\tau)=\frac{c\gamma}{4}\tau^{\gamma-1},
\label{Ktau}
\end{equation}
\begin{equation}
\eta (\Delta l)=\frac{c\gamma}{4}((\Delta l)^{2}/c)^{(\gamma-1)/\gamma}.
\label{K_l}
\end{equation} 

As if follows from Figure \ref{fig6}, for both data sets we observe the
super-diffusion regime. The coefficient of magnetic turbulent diffusivity,
$\eta(\Delta l)$, derived by Eq. (\ref{K_l}) for both data sets is shown in
Figure \ref{fig6}{\it b}. Two essential things should be mentioned here: first,
the diffusion coefficient grows as the scale increases (the same is true for a
time scale, too, see Eq. (\ref{Ktau}). Second, the slope of the power law varies
with scale (which is a characteristic feature of intrinsic multi-fractality). On
the minimal spatial (25 km) and temporal (10 sec) scales considered in
\citet{Abramenko+2011-Diff},
 the diffusion coefficient in
QS area was found to be 19~km$^{2}$~s$^{-1}$. The HMI data provided a value of
approximately 220~km$^{2}$~s$^{-1}$ on the largest available scale of 4 Mm.

The observed tendency of the turbulent magnetic diffusivity to
decrease with decreasing scales leads us to expect that the turbulent
diffusivity might be close to the magnitudes of diffusivity adopted in the
numerical
simulations of small-scale dynamo (0.01 - 10 km$^2$s$^{-1}$, e.g., 
\citet{boldyrev2004,Vogler-Schussler-2007-SSD,PietarilaGraham_2010}).
This makes the simulations even more realistic.

In summary, a super-diffusion regime on very small scales is
very favorable for pictures assuming turbulent dynamo action since it assumes
decreasing diffusivity with decreasing scales.

\section{Concluding remarks}
Continuously varying magnetic fields are the main reason for the solar/stellar
activity. The 11-year solar cycle is one of the most astonishing and widely
known examples of the self-organized generation of the magnetic field. Although
we know that there is no two absolutely similar solar cycles, yet, persistency
and regularity of the solar periodicity through thousands of years remains
impressive. A drastically different picture arises when one looks on the
photosphere: chaos of mixed-polarity magnetic elements of all sizes until the
resolution limits of modern instruments, continuously renewing during 1-2 days -
the magnetic carpet. 

Dualism of the solar magnetism is usually explained by a simultaneous action of
two dynamos: a global dynamo operating in the convective zone and responsible
for the 11-year solar cycle, and local, or turbulent dynamo, which might operate
inside the near-surface layer and to be responsible for generation of
small-scale magnetic fields forming the magnetic carpet. The explanation seems
to oversimplify the reality because resent studies of distribution of the
magnetic flux accumulated in magnetic flux tubes showed the non-interrupted
power law for many decades 
\citep{Parnell_2009} thus
supposing a common (for all scales) mechanism for the magnetic field generation.
One of promising ways to handle the problem is to consider the solar dynamo
process as a non-linear dynamical system (NDS), with intrinsic properties of
multi-fractality and intermittency.

Like any NDS, the solar dynamo is then capable to self-organization on all
scales (including large scales) and display a chaotic nature on small scales.
Self-organization, in turn, provides for a magnetic complex a way to reach a SOC
state, when burst-like energy release events of any size are possible at any
time instant. The concept is very important for our understanding of flaring and
heating processes in solar/stellar atmospheres.

Further, multi-fractal nature on the magnetic field provides a
necessary condition for the local turbulent dynamo operation in the
near-surface layer of the convective zone. Observational evidences for local
dynamo operation are still under strong debates, e.g., compare
the talks by Drs. Tsuneta and Stenflo presented at this symposium.

One pragmatic advise for researchers could be inferred form the observed
multi-fractal nature of magnetized solar plasma. Namely, observed power laws
should not be extrapolated over neighboring scales, a frequent mistake for power
laws studies in various fields. 

In summary, the paradigm of multi-fractal and highly intermittent structure of
solar magnetized plasma offers new approaches to understand the solar and
stellar magnetism.

%\bibliographystyle{$HOME/texmf/tex/latex/aastex/apj}

%\bibliography{$HOME/abramenko}

\end{document}